\documentclass[8.5pt,twoside,twocolumn]{article}
\usepackage[super,sort&compress,comma]{natbib}
\usepackage{mhchem}
\usepackage{times}
\usepackage{sectsty}
\usepackage{balance}

\usepackage{graphicx} 
\usepackage{lastpage}
\usepackage[format=plain,justification=raggedright,singlelinecheck=false,font=small,labelfont=bf,labelsep=space]{caption}
\usepackage{fancyhdr}
\begin{document}




\twocolumn[
  \begin{@twocolumnfalse}
\noindent\LARGE{\textbf{Long-range interactions between ultracold atoms and molecules including atomic spin-orbit}}
\vspace{0.6cm}

\noindent\large{\textbf{Maxence Lepers $^{\ast}$\textit{$^{a}$} and Olivier Dulieu\textit{$^{a}$}}}
\vspace{0.5cm}

\noindent\textit{\small{\textbf{Received Xth XXXXXXXXXX 20XX, Accepted Xth XXXXXXXXX 20XX\newline
First published on the web Xth XXXXXXXXXX 200X}}}

\noindent \textbf{\small{DOI: 10.1039/b000000x}}
\vspace{0.6cm}

\noindent \normalsize{We investigate theoretically the long-range electrostatic interactions between a ground-state homonuclear
alkali-metal dimer and an excited alkali-metal atom taking into account its fine-structure. The interaction involves the combination of first-order quadrupole-quadrupole and second-order dipole-dipole effects. Depending on the considered species, the atomic spin-orbit may be comparable to the atom-molecule electrostatic energy and to the dimer rotational structure. Here we extend our general description in the framework of the second-order degenerate perturbation theory [M. Lepers and O. Dulieu, Eur. Phys. J. D, 2011] to various regimes induced by the magnitude of the atomic spin-orbit. A complex dynamics of the atom-molecule may take place at large distances, which may have consequences for the search for an universal model of ultracold inelastic collisions as proposed for instance in [Z. Idziaszek and P. S. Julienne, Phys. Rev. Lett. \textbf{104}, 113202 (2010)]. }
\vspace{0.5cm}
 \end{@twocolumnfalse}
  ]



\footnotetext{\textit{$^{a}$~Laboratoire Aim\'e Cotton, UPR3321 CNRS, B\^at.~505, Univ Paris-Sud, 91405 Orsay Cedex, France Fax: 33 16941 0156; Tel: 33 16935 2013; E-mail: maxence.lepers@lac.u-psud.fr}}



\section{Introduction}
\label{sec:intro}

As illustrated by the present issue of the journal \cite{pccp_edito2011}, the field of cold and ultracold molecules is continuously developing in many directions of fundamental and interdisciplinary physics. Among many developments, researchers can now create large samples of ultracold molecules which can undergo elastic, inelastic, and reactive collisions with surrounding ultracold atoms \cite{elioff2003,zahzam2006,staanum2006,hudson2008}. The challenge is at least twofold: the achievement of sympathetic cooling of molecules down to quantum degeneracy \cite{cybulski2005,tacconi2007,lara2007,tscherbul2010}, and the control of elementary chemical reaction at cold and ultracold temperatures \cite{willitsch2008,tscherbul2008,levinsen2009,ospelkaus2010a,knoop2010}. As pointed out by Julienne (see for instance Ref.\cite{julienne2009}), low-energy atom-molecule inelastic collisions can be understood as resulting from the dynamics at large distances controlled by long-ranges forces, combined to the dynamics induced by short-range chemical forces. The resulting rate can be written as a product involving scattering probabilities in both domains. Due to their complex internal structure, molecules most often offer many open channels for inelastic or reactive processes with other systems so that the related probability can be assumed equal to one, so that the entire collision is uniquely controlled by well-known long-range forces. Just like for ultracold elastic collisions determined by a single parameter, namely the scattering length, this approach opens the way for elaborating universal models for inelastic and reactive atom-molecule collisions, as recently proposed by several authors \cite{idziaszek2010a,idziaszek2010b,gao2010,micheli2010}. In such models, it is assumed that the inelastic rates depend solely on the leading term $C_n/R^n$ of the long-range electrostatic interaction between the colliding partners.

Most of the cases investigated in the papers above involve atoms and molecules in their electronic ground state. In a series of recent papers we studied the long-range interaction between an alkali-metal dimer in its ground state with an excited alkali-metal atom, in the perspective of modeling their association into an excited trimer induced by a properly chosen laser (photoassociation, or PA) at ultracold energies. In Refs. \cite{lepers2010,lepers2011a} (hereafter referred to as Papers I and II, respectively) we have characterized the first-order quadrupole-quadrupole and second-order dipole-dipole interactions -- varying as $C_5/R^5$ and $C_6/R^6$ (where $R$ is the distance between the colliding partners) and associated with the operators $\hat{V}_{qq}$ and $\hat{V}_{dd}^{(2)}$, respectively -- between a homonuclear alkali-metal dimer in the lowest level $v_d=0$ of its electronic ground state $X^1\Sigma_g^+$, and an alkali-metal atom in the first excited state $^2P$. We have located a range of atom-dimer distances with an upper bound $R_{p}^{*}$ inside which both kinds of interaction compete with the rotational energy of the dimer. Potential energy curves asymptotically connected to different dimer rotational levels $N$ are thus coupled, inducing complex patterns further characterized in Ref.\cite{lepers2011b} (hereafter referred to as Paper III). The lower bound of this region is limited by the so-called LeRoy radius \cite{leroy1974} below which exchange interaction takes place.

In this series of three papers we have applied our formalism to the interaction between a Cs$_2$ dimer and a Cs atom but it can be applied
to all alkali-metal combinations. Considering other species will modify the limits of the $\left[ R_{LR};R_{p}^* \right]$ region but not the overall aspect of the potential energy curves. This is due to the moderate variation of the electrostatic and rotational energies from Li to Cs (see Table \ref{tab:ODG-ener}). In contrast, the spin-orbit splitting of the first excited atomic state $n^2P$ varies dramatically from Li to Cs. Lithium is actually the only species whose fine-structure splitting is comparable to the rotational and electrostatic energies. For all the other species, the spin-orbit splitting is much higher. Following \cite{dubernet1994}, we define two distinct coupling cases in analogy to Hund's cases:
\begin{itemize}
\item From Na to Cs, the fine-structure splitting is so large that the two fine-structure components are not coupled by the electrostatic interaction. The spin-orbit Hamiltonian $\hat{H}_{SO}$ is part of the zeroth-order energy $E_{p,1}^{0}=E_{X,v_d=0}+E_{nP} + \left\langle\hat{H}_{SO}\right\rangle$ of a given state $p$, where $E_{X,v_d=0}$ is the energy of the electronic and vibrational dimer ground state, and $E_{nP}$ the energy of the atomic $n^2P$ state without fine structure. The electrostatic interaction is then obtained by diagonalizing at each $R$ the Hamiltonian
\begin{equation}
\hat{W}_1(R) = B_0\hat{\vec{N}}^2 + \hat{V}_{qq} + \hat{V}_{dd}^{(2)}\label{eq:W1} \,,
\end{equation}
with $B_0$ the dimer rotational constant in the electronic and vibrational ground state. This situation, denoted as case ``{}1C" in Ref.\cite{dubernet1994}, will be illustrated with cesium in the $6^{2}P$ manifold.
\item For Li in the $2^{2}P$ manifold, the two fine-structure components are coupled by the electrostatic interaction, and the spin-orbit interaction is included in the perturbation Hamiltonian
\begin{equation}
\hat{W}_2(R) = B_0\hat{\vec{N}}^2 + \hat{H}_{SO} + \hat{V}_{qq} + \hat{V}_{dd}^{(2)}\,.
\label{eq:W2}
\end{equation}
The zeroth-order energy of state $p$ reads $E_{p,2}^{0}=E_{X,v_d=0}+E_{nP}$. Due to the very weak variation of the electrostatic properties between the $2^{2}P_{1/2}$ and $2^{2}P_{3/2}$ states, we will consider that those properties do not depend on the fine-structure level. This situation corresponds to case ``{}2A" in Ref.\cite{dubernet1994}.
\end{itemize}
\begin{table}[h]
\small
\centering
  \caption{\label{tab:ODG-ener}\ Some atomic properties relevant for the present study: the mean square radius $\left\langle r_{np}^{2}\right\rangle$ on the atomic lowest $np$ orbital, its spin-orbit splitting $\Delta E_{fs}$, the rotational constant $B_0$ of the lowest vibrational level of the corresponding dimer, and the only nonzero tensor component $q_2^0$ of the dimer quadrupole moment at its equilibrium distance (see Paper I)}
  \begin{tabular}[width=0.5\textwidth]{@{\extracolsep{\fill}}cccc}
    \hline
    & Li & K & Cs \\
    $\left\langle r_{np}^{2}\right\rangle$ (a.u.) &27.1   & -      & 62.7 \\
    $\Delta E_{fs}$ (cm$^{-1}$)                   & 0.335 & 57.7   & 554 \\
    \hline
                                                  &Li$_2$ &K$_2$   &Cs$_2$ \\
    $q_2^0$ (a.u.)                                &10.5   &15.7    & 18.6 \\
    $B_0$ (cm$^{-1}$)                             & 0.673 & 0.0567 & 0.0117 \\
    \hline
  \end{tabular}
\end{table}
In Section \ref{sec:general} we briefly recall the general formalism that we used in our previous work, emphasizing on the modifications induced by the presence of the atomic spin-orbit. We derive in Section \ref{sec:longrange} the expressions for the first-order and second-order long-range interaction of the atom-molecule system, referring to the two cases ``{}1C" and ``{}2A" above. Then we describe our results for the Cs$^*$+Cs$_2$ case (Section \ref{sec:Cs-Cs2}) and for the Li$^*$+A$_2$ case, A being an alkali-metal species (Section \ref{sec:Li-A2}). Finally we discuss the implications of our results in the perspective of the universal model proposed in Ref.\cite{julienne2009,idziaszek2010a} (Section \ref{sec:discussion}).

\section{General formalism}
\label{sec:general}

\begin{figure}[h]
\centering
  \includegraphics[width=0.4\textwidth]{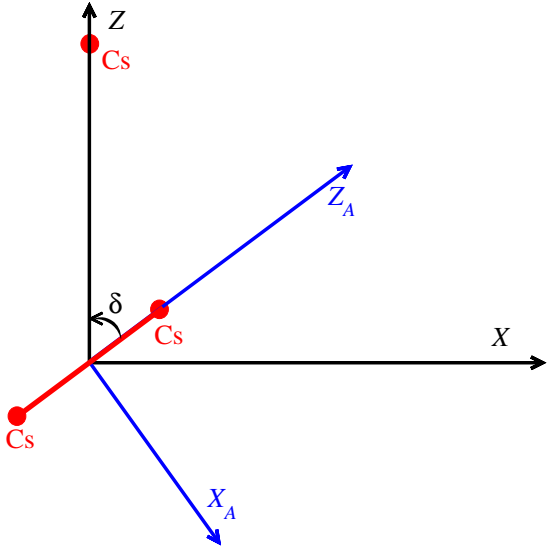}
  \caption{(Color online) The two coordinate systems, $X_{A}Y_{A}Z_{A}$ (D-CS) and $XYZ$ (T-CS) defined for the dimer and for the trimer, respectively. The $Z_{A}$ axis is along the dimer axis, while $Z$ is oriented from the center of mass of the dimer towards the atom $B$. The $Y$ and $Y_{A}$ axes coincide and point into the plane of the figure. The subsystem $A$ in this figure is the Cs$_{2}$ molecule, the subsystem $B$ is the Cs atom. The T-CS is related to the laboratory coordinate system ($\tilde{x}\tilde{y}\tilde{z}$) by the usual Euler angles ($\alpha,\beta,\gamma$), not represented here.}
  \label{fig:CS}
\end{figure}

We start from the general form of the electrostatic energy between two interacting charge distributions $A$ and $B$, whose centers of mass are separated by the distance $R$,
\begin{eqnarray}
\hat{V}_{el}(R) & = & \sum_{L_{A},L_{B}=0}^{+\infty}\sum_{M=-L_{<}}^{L_{<}}\frac{1}{R^{1+L_{A}+L_{B}}}\nonumber \\
 & \times & f_{L_{A}L_{B}M}\hat{Q}_{L_{A}}^{M}(\hat{r}_{A})\hat{Q}_{L_{B}}^{-M}(\hat{r}_{B})\,,
\label{eq:MultipoleExp}
\end{eqnarray}
where $L_{<}$ is the minimum of $L_{A}$ and $L_{B}$. Eq.~(\ref{eq:MultipoleExp}) is the well-known expansion on the electric multipoles of $A$ and $B$. Each multipole of order $L_{X}$ (with $X=A,B$) is associated with the tensor operator $\hat{Q}_{L_{X}}^{M}(\hat{r}_{X})$, which can be expressed in a coordinate system whose origin is the center of mass of $X$
\begin{equation}
\hat{Q}_{L_{X}}^{M}(\hat{r}_{X})=\sqrt{\frac{4\pi}{2L_{X}+1}}\sum_{i\in X}q_{i}\hat{r}_{i}^{L_{X}}Y_{L_{X}}^{M}\left(\hat{\theta}_{i},\hat{\phi}_{i}\right)\,,
\label{eq:OperMultip}
\end{equation}
where $q_{i}$ is the value of each charge $i$ composing $X$, and $Y_{L_{X}}^{M}$ are the usual spherical harmonics. In Eq. (\ref{eq:MultipoleExp}), the assumption has be made that the quantization axis is the one pointing from the center of mass of $A$ to the center of mass of $B$, hence the factor $f_{L_{A}L_{B}M}$ reads
\begin{eqnarray}
f_{L_{A}L_{B}M} & = & \frac{\left(-1\right)^{L_{B}}\left(L_{A}+L_{B}\right)!}{\sqrt{\left(L_{A}+M\right)!\left(L_{A}-M\right)!}} \nonumber\\
 & \times & \frac{1}{\sqrt{\left(L_{B}+M\right)!\left(L_{B}-M\right)!}}.
\end{eqnarray}
In our particular case (Fig. \ref{fig:CS}), the system $A$ is the dimer, the system $B$ is the atom, and the quantization axis $Z$ joins the dimer center of mass and the atom.

In what follows, we want to characterize the first-order quadrupole-quadrupole interaction (defined by $L_A=L_B=2$ in Eq.~(\ref{eq:MultipoleExp})) and the second-order dipole-dipole interaction, (defined by $L_A=L_B=1$), in the jj coupling case. Namely, the atomic state is characterized by the quantum numbers associated with the outermost electron: $n$ the radial quantum number, $\ell$ the orbital angular momentum, $j$ the total angular momentum, and $\omega$ its projection on the Z axis, joining the center of mass of the dimer and the atom. We shall also use $s$ the spin of the electron ($s=1/2$), its projection $\sigma$ on the Z axis, as well as $\lambda$, the projection of $\ell$. Using these notations, the atomic spin-orbit Hamiltonian reads
\begin{equation}
\hat{H}_{SO} = \mathcal{A}\hat{\vec{\ell}}.\hat{\vec{s}} \,,
\end{equation}
where $\mathcal{A}=2\Delta E_{fs}/3$, and its matrix elements are diagonal
\begin{equation}
\left\langle\hat{H}_{SO}\right\rangle = \frac{\mathcal{A}}{2}\left(j(j+1)-\ell(\ell+1)-s(s+1)\right) \,.
\end{equation}
The diatomic molecule is in its ground electronic state $\left|X^{1}\Sigma_{g}^{+}\right\rangle $ and vibrational level $\left|v_{d}=0\right\rangle $. The quantum numbers associated with the rotation of the dimer nuclei are the angular momentum $N$ and its projection $m$ on the Z axis. The total angular momentum $J$, associated with the mutual atom-dimer rotation, will be considered in a future work. On the contrary, the projection
\begin{equation}
\Omega=m+\omega
\label{eq:Omega}
\end{equation}
of $\vec{J}$ on the Z axis, will be extensively used, as it is a conserved quantity. In comparison to Paper III, the atomic part of the electrostatic interaction is the only one which is modified. That is why we will focus on it in what follows.

The atomic multipole moments have now to be expressed in the $jj$ coupling case. As the electric-multipole-moment operators $\hat{Q}_{L}^{M}$ act on orbital part of the atomic state, we first decompose the atomicstate, labeled $\left|j\omega\right\rangle $ on the corresponding decoupled basis $\left\{ \left|\lambda\sigma\right\rangle \right\}
$\begin{equation}
\left|j\omega\right\rangle =\sum_{\lambda\sigma}C_{\ell\lambda s\sigma}^{j\omega}\left|\lambda\sigma\right\rangle ,
\end{equation}
where $C_{a\alpha b\beta}^{c\gamma}$ is a Clebsch-Gordan coefficient. Combining the matrix element of $\hat{Q}_{L}^{M}$ in the decoupled basis,
\begin{eqnarray}
 & & \left\langle n_{1}\ell_{1}\lambda_{1}\sigma_{1}\right|\hat{Q}_{L}^{M}\left|n_{2}\ell_{2}\lambda_{2}\sigma_{2}\right\rangle \nonumber\\
 & = & -\delta_{\sigma_{1}\sigma_{2}}\sqrt{\frac{2\ell_2+1}{2\ell_1+1}}\left\langle \hat{r}_{12}^{L}\right\rangle C_{\ell_{2}0L0}^{\ell_{1}0} C_{\ell_{2}\lambda_{2}LM}^{\ell_{1}\lambda_{1}} \,,
\end{eqnarray}
where $\left\langle \hat{r}_{12}^{L}\right\rangle \equiv\left\langle \hat{r}_{n_{1}\ell_{1}j_{1}n_{2}\ell_{2}j_{2}}^{L}\right\rangle $ is the matrix element associated with the operator $\hat{r}^{L}$ for the outermost electron of the atom, and the formula \cite{varshalovich1988}
\begin{eqnarray}
\sum_{\alpha\beta\delta}C_{a\alpha b\beta}^{c\gamma}C_{d\delta b\beta}^{e\epsilon}C_{a\alpha f\phi}^{d\delta} & = & \left(-1\right)^{b+c+d+f}\sqrt{\left(2c+1\right)\left(2d+1\right)}\nonumber \\
 & \times & \left\{ \begin{array}{ccc}
a & b & c\\
e & f & d\mbox{}\end{array}\right\} C_{c\gamma f\phi}^{e\epsilon}\,,
\label{eq:prod-3Clebsch}\end{eqnarray}
where $\left\{ \begin{array}{ccc}
a & b & c\\
d & e & f\end{array}\right\} $ is a Wigner 6-j symbol, and using the fact that $\ell_1+L$ has the same parity as $\ell_2$, we get to the expression
\begin{eqnarray}
 & & \left\langle n_{1}\ell_{1}j_{1}\omega_{1}\right|\hat{Q}_{L}^{M}\left|n_{2}\ell_{2}j_{2}\omega_{2}\right\rangle \nonumber\\
 & = & \left(-1\right)^{\ell_{2}+s+j_{2}+1}\sqrt{\left(2\ell_{2}+1\right)\left(2j_{2}+1\right)}\left\langle \hat{r}_{12}^{L}\right\rangle \nonumber \\
 & \times & C_{\ell_{2}0L0}^{\ell_{1}0}
 \left\{ \begin{array}{ccc}
\ell_{2} & s & j_{2}\\
j_{1} & L & \ell_{1}
\end{array}\right\}
C_{j_{2}\omega_{2}LM}^{j_{1}\omega_{1}}\,.
\label{eq:QLM-gene}
\end{eqnarray}
In case ``{}2A" defined in the previous section, the radial matrix element $\hat{r}^{L}$ is independent from $j_1$ and $j_2$.

\section{First-order and second-order long-range interactions}
\label{sec:longrange}

The matrix element corresponding to the quadrupolar interaction $\hat{V}_{qq}$ is obtained by combining Eq. (\ref{eq:QLM-gene}) with $L=2$ and $\ell_1=\ell_2$ for the atomic part, and Eq.~(12) of Paper III for the dimer part, which yields
\begin{eqnarray}
 & & \left\langle N_{1}m_{1}j_{1}\omega_{1}\left|\hat{V}_{qq}\right|N_{2}m_{2}j_{2}\omega_{2}\right\rangle \nonumber \\
 & = & 24\left(-1\right)^{\ell+j_{2}+3/2}\sqrt{\frac{2N_{2}+1}{2N_{1}+1}}\sqrt{\left(2\ell_2+1\right)\left(2j_{2}+1\right)}\nonumber \\
 & \times & C_{N_{2}020}^{N_{1}0}C_{\ell020}^{\ell0}
 \left\{ \begin{array}{ccc}
\ell & \frac{1}{2} & j_{2}\\
j_{1} & 2 & \ell\end{array}\right\}
\frac{q_{2}^{0}\left\langle r_{n\ell j_{1}n\ell j_{2}}^{2}\right\rangle }{R^{5}}\nonumber \\
 & \times & \sum_{M=-2}^{2}\frac{C_{N_{2}m_{2}2M}^{N_{1}m_{1}}C_{j_{2}\omega_{2}2-M}^{j_{1}\omega_{1}}}{\left(2+M\right)!\left(2-M\right)!}\,,
\label{eq:Vqq}
\end{eqnarray}
where $q_{2}^{0}$ is the tensor component of the dimer quadrupole moment along its internuclear axis $Z_A$. The angular factors of Eq.~(\ref{eq:Vqq}) impose strong selection rules: (i) $m_1+\omega_1=m_2+\omega_2$, which means that the quantum number $\Omega$ (see Eq.~(\ref{eq:Omega})) is conserved; (ii) $N_{1}=N_{2},\: N_{2}\pm2$; and (iii) $j_{1}$, $j_{2}$ and $L=2$ must satisfy the triangle rule. It is important to remark that the latter selection rule is \textit{not} satisfied for $j_{1}=j_{2}=1/2$, which is of strong importance for the Cs$_{2}+$Cs interaction (case ``{}1C"). If the cesium atom is in the $6^{2}P_{1/2}$ fine-structure level, it only interacts with Cs$_{2}$ through the second-order dipolar interaction, which is certainly not favorable for the existence of long-range vibrational levels below the $6^{2}P_{1/2}$ dissociation limit of the trimer.

The second-order dipolar interaction is associated with the operator $\hat{V}_{dd}^{(2)}$, whose matrix elements can be written as functions of the dynamical polarizabilities of the two fragments at imaginary frequencies (see Eq. (12) of Paper II)
\begin{eqnarray}
 & & \left\langle N_{1}m_{1}j_{1}\omega_{1}\left|\hat{V}_{dd}^{(2)}\right|N_{2}m_{2}j_{2}\omega_{2}\right\rangle \nonumber \\
 & = &
-\sum_{M=-1}^{1}\sum_{M'=-1}^{1}\frac{4}{\left(1+M\right)!\left(1-M\right)!\left(1+M'\right)!\left(1-M'\right)!}\nonumber \\
 & \times
& \left[\frac{1}{2\pi}\int_{0}^{+\infty}d\omega \alpha_{MM'}^{m_{1}m_{2}}(i\omega)\alpha_{-M-M'}^{\omega_{1}\omega_{2}}(i\omega)\right.\nonumber \\
 &  &
\left.+\sum_{b}\Theta(-\Delta E_{b}^{0})\alpha_{MM'}^{m_{1}m_{2}}(\omega=\Delta E_{b}^{0})\right.\nonumber \\
 &  &
\left.\times\left\langle n\ell j_{1}\omega_{1}\left|\hat{Q}_{1}^{-M} \right|\Phi_{b}^{0}\right\rangle
\left\langle \Phi_{b}^{0}\left|\hat{Q}_{1}^{M'}\right|n\ell j_{2}\omega_{2}\right\rangle \right]\,,
\label{eq:Vdd2-1}
\end{eqnarray}
where $\Theta(x)$ is Heaviside function, and the letter $b\equiv n'\ell'j'$ stands for all the quantum states of the atom accessible through dipolar transitions. The last two lines of Eq.~(\ref{eq:Vdd2-1}) are contributions due to the downward atomic transitions. They depend on the dimer dynamical polarizability $\alpha_{MM'}^{m_{1}m_{2}}$ at the (real) frequencies of the atomic transitions, which is given in Eq. (20) of Paper III.

In Eq.~(\ref{eq:Vdd2-1}), we have introduced the atomic dynamical polarizability in the coupled basis
\begin{eqnarray}
 & & \alpha_{-M-M'}^{\omega_{1}\omega_{2}}(z) \nonumber\\
 & = & 2\left(-1\right)^{M}\sum_{b}\frac{\left(E_{b}-E_{n\ell{j_1}}\right)}{\left(E_{b}-E_{n\ell{j_1}}\right)^{2}-z^{2}}\nonumber \\
 &  & \times\left\langle n\ell{j_1}\omega_{1}\left|\hat{Q}_{1}^{-M}\right|\Phi_{b}\right\rangle \left\langle \Phi_{b}\left|\hat{Q}_{1}^{M'}\right|n\ell{j_2}\omega_{2}\right\rangle,
\label{eq:alpha-1}
\end{eqnarray}
where $z$ can be either real or complex. The first line of Eq.~(\ref{eq:alpha-1}) depends on $j_1$, but not on $j_2$: the subtle reason for this will be explained in the what follows. By applying Eq.~(\ref{eq:QLM-gene}) to $L=1$, and putting the primes in upper indices of the Clebsch-Gordan coefficients using the identity
\begin{equation}
C_{a\alpha b\beta}^{c\gamma}=\left(-1\right)^{a+2b-c-\beta}\sqrt{\frac{2c+1}{2a+1}}C_{c\gamma b-\beta}^{a\alpha}\,,
\end{equation}
we get to the final expression
\begin{eqnarray}
 & & \alpha_{-M-M'}^{\omega_{1}\omega_{2}}(z) \nonumber\\
 & = & 2\sum_{n'\ell'j'\omega'}\frac{\left(E_{n'\ell'j'}-E_{n\ell{j_1}}\right)}{\left(E_{n'\ell'j'}-E_{n\ell{j_1}}\right)^{2}-z^{2}}\nonumber \\
 &  & \times \left(-1\right)^{1+j_1+j_2}
\left(2\ell+1\right)\sqrt{\left(2j_{1}+1\right)\left(2j_{2}+1\right)}  \nonumber \\
 &  & \times\left\langle \hat{r}_{n\ell j_{1}n'\ell'j'}\right\rangle \left\langle \hat{r}_{n\ell j_{2}n'\ell'j'}\right\rangle \left(C_{\ell010}^{\ell'0}\right)^{2}\nonumber \\
 &  & \times\left\{ \begin{array}{ccc}
\ell & \frac{1}{2} & j_{1}\\
j' & 1 & \ell'\end{array}\right\} \left\{ \begin{array}{ccc}
\ell & \frac{1}{2} & j_{2}\\
j' & 1 & \ell'\end{array}\right\} \nonumber\\
 & & \times C_{j_{1}\omega_{1}1M}^{j'\omega'} C_{j_{2}\omega_{2}1M'}^{j'\omega'}\,.
\label{eq:alpha-2}\end{eqnarray}
The key point is now to connect Eq.~(\ref{eq:alpha-2}) to the isotropic polarizability of the atom, which will depend on the considered cases ``{}1C" and ``{}2A".

In the case ``{}1C" illustrated by cesium, the isotropic polarizabilities are different for the two fine-structure levels $6^{2}P_{1/2}$ and $6^{2}P_{3/2}$. Moreover, as the subspaces associated to those two levels are fully decoupled, the matrix element of $\hat{V}_{dd}^{(2)}$ are zero if $j_{1}\neq j_{2}$. As pointed out in Paper II, the isotropic polarizability $\bar{\alpha}{}_{n\ell j}$ corresponding to the level $n\ell_{j}$ with sublevels $\omega$ (not to be mixed up with the frequency) is obtained from Eq.~(\ref{eq:alpha-2}) by carrying out a sum over all values of $\omega'$ and an average over $\omega$
\begin{eqnarray}
{\bar{\alpha}}_{n\ell j}(z) & = & \alpha_{c}+\frac{2}{3}\sum_{n'\ell'j'}\frac{\left(E_{n'\ell'j'}-E_{n\ell j}\right)}{\left(E_{n'\ell'j'}-E_{n\ell j}\right)^{2}-z^{2}}\nonumber \\
 &  & \times\left(2\ell+1\right)\left(2j'+1\right)\left\langle \hat{r}_{n\ell jn'\ell'j'}\right\rangle ^{2}\nonumber \\
 &  & \times\left(C_{\ell010}^{\ell'0}\right)^{2}
 \left\{ \begin{array}{ccc}
\ell & \frac{1}{2} & j\\
j' & 1 & \ell'\end{array}\right\} ^{2}\,.
\end{eqnarray}
Here $\alpha_{c}$ is the polarizability of the atomic core which is the same as in Paper II, and we used the identity $\sum_{\alpha\gamma}\left(C_{a\alpha b\beta}^{c\gamma}\right)^{2}=\frac{2c+1}{2b+1}$. The situation is thus similar to the one of Paper II: because of angular factors, the quantity $\alpha_{-M-M'}^{\omega_{1}\omega_{2}}$ cannot be related to the sole polarizability of the $n\ell_{j}$ atomic level, and contributions from different $j\to j'$ transitions must be separated. For example, the contributions $j=3/2\,\to j'=1/2$, $j=3/2\,\to j'=3/2$ and $j=3/2\,\to j'=5/2$ for cesium in the $6^{2}P_{3/2}$ state must be distinguished. Following Eq.~(24) of Paper II, we express the isotropic polarizability as
\begin{equation}
\bar{\alpha}_{n\ell j}=\sum_{n'\ell'j'}\alpha_{n\ell jn'\ell'j'}+\alpha_{c}\,,
\end{equation}
 where the state-to-state polarizability $\alpha_{n\ell jn'\ell'j'}$ is related to $\alpha_{-M-M'}^{\omega_{1}\omega_{2}}$ by
\begin{eqnarray}
\alpha_{-M-M'}^{\omega_{1}\omega_{2}} & = & 3\delta_{j_1j_2}\sum_{j'=j_1-1}^{j_1+1}\frac{2j_1+1}{2j'+1}\nonumber \\
 & \times & \sum_{\ell'=(\ell-1,\ell+1)}\sum_{n'}\alpha_{n\ell j_1n'\ell'j'}\nonumber \\
 & \times & \sum_{\omega'=-j'}^{+j'}C_{j_{1}\omega_{1}1M}^{j'\omega'}C_{j_{1}\omega_{2}1M'}^{j'\omega'}\,.\end{eqnarray}
It is important to keep in mind that the quantum numbers $\ell'$ and $j'$ are related to each other by the condition $\ell'-s\le j'\le\ell'+s$. So, the possible transitions are: $n^{2}P_{1/2}\to n'^{2}S_{1/2}$ and $n^{2}P_{1/2}\to n'^{2}D_{3/2}$, for an alkali-metal atom in a $n^{2}P_{1/2}$ state, and $n^{2}P_{3/2}\to n'^{2}S_{1/2}$, $n^{2}P_{3/2}\to n'^{2}D_{3/2}$ and $n^{2}P_{3/2}\to n'^{2}D_{5/2}$, for an alkali-metal atom in a $n^{2}P_{3/2}$ state.

The case "1A" corresponds to a lithium atom in the $2^{2}P$ state. As we have not calculated the dynamical polarizability of lithium, we will only present the formalism and no numerical results. We assume that the different fine-structure levels of the $n\ell$ manifold can be coupled by electrostatic interaction, but that the energies of the states $n\ell_{j}$ and $n'\ell'_{j'}$, and the transition dipole moments from state $n\ell_{j}$ to state $n'\ell'_{j'}$ do not depend on $j$ and $j'$. It is thus more relevant to express the isotropic polarizability $\bar{\alpha}_{n\ell}$ in the Russel-Sanders coupling scheme, which is given in Paper II, Eqs. (22) znd (24),
\begin{eqnarray}
{\bar{\alpha}}_{n\ell}(z) & = & \alpha_{c}+\frac{2}{3}\sum_{n'\ell'}\frac{\left(E_{n'\ell'}-E_{n\ell}\right)}{\left(E_{n'\ell'}-E_{n\ell}\right)^{2}-z^{2}}\nonumber \\
 &  & \times\left\langle \hat{r}_{n\ell n'\ell'}\right\rangle ^{2}\left(C_{\ell010}^{\ell'0}\right)^{2}.\end{eqnarray}
In order to calculate $\alpha_{MM'}^{\omega_{1}\omega_{2}}(z)$, it is, like in Paper II, necessary to separate the different $\ell\to\ell'$
transitions (here $P\to S$ and $P\to D$). By introducing the state-to-state polarizability,
\begin{eqnarray}
\alpha_{n\ell n'\ell'}(z) & = & \frac{2}{3}\frac{\left(E_{n'\ell'}-E_{n\ell}\right)}{\left(E_{n'\ell'}-E_{n\ell}\right)^{2}-z^{2}}\nonumber \\
 &  & \times\left\langle \hat{r}_{n\ell n'\ell'}\right\rangle ^{2}\left(C_{\ell010}^{\ell'0}\right)^{2},
\end{eqnarray}
we finally get to the relation
\begin{eqnarray}
\alpha_{-M-M'}^{\omega_{1}\omega_{2}} & = & 3\sum_{\ell'=(\ell-1,\ell+1)}\sum_{n'}\alpha_{n\ell n'\ell'}\nonumber \\
 & \times & \sum_{j'=j_>-1}^{j_<+1}\left(2\ell+1\right)\sqrt{\left(2j_{1}+1\right)\left(2j_{2}+1\right)}\nonumber \\
 & \times & \left\{ \begin{array}{ccc}
\ell & \frac{1}{2} & j_{1}\\
j' & 1 & \ell'\end{array}\right\} \left\{ \begin{array}{ccc}
\ell & \frac{1}{2} & j_{2}\\
j' & 1 & \ell'\end{array}\right\} \nonumber \\
 & \times & \sum_{\omega'=-j'}^{+j'}C_{j_{1}\omega_{1}1M}^{j'\omega'} C_{j_{2}\omega_{2}1M'}^{j'\omega'}\,,\end{eqnarray}
with $j_>$ the maximum of $j_1$ and $j_2$, and $j_<$ their minimum.

\section{Case ``{}1C": Cs$^*$+Cs$_{2}$}
\label{sec:Cs-Cs2}

In this case the two atomic fine-structure subspaces are fully decoupled and treated separately. Figs.~\ref{fig:Cs2-Cs-j32-Om12} and \ref{fig:Cs2-Cs-j32-Om32} present the potential energy curves characterizing the interaction between a Cs$_2$ dimer in the lowest rotational levels $N$ and a Cs atom in the $6^2P_{1/2}$ level, obtained after diagonalization of $\hat{W}_1(R)$ (see Eq.~(\ref{eq:W1})). The curves are sorted by values of $|\Omega|$ (here equal to 1/2 and 3/2) and parity of $N$. Table \ref{tab:C6-j12} presents the corresponding $C_6$ coefficients for $N=0$ and 1, all the $C_5$ coefficients being zero. The method used to calculate the polarizabilities of Cs$_2$ and Cs are the same as those described in Paper II.

\begin{figure}
\centering
\includegraphics[width=0.4\textwidth]{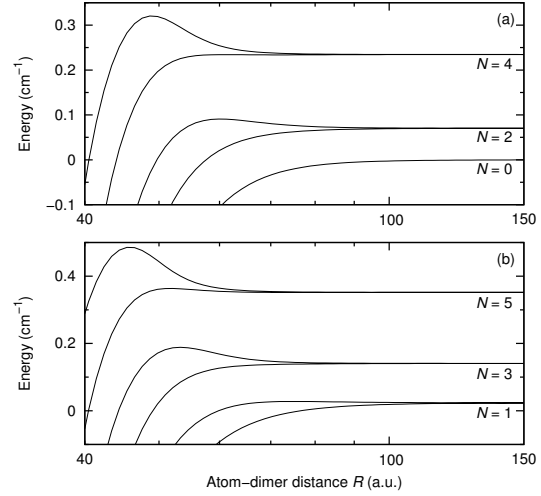}
\caption{\label{fig:Cs2-Cs-j12-Om12}Long-range potential energy curves between a ground-state Cs$_{2}$ and an excited Cs($6^{2}P_{j=1/2}$) atom, as functions of their mutual separation $R$, for $|\Omega|=1/2$ and for: (a) the even values of $N$ and (b) the odd values of $N$. The origin of energies is taken at the Cs$_2(X,v_d=0,N=0)+$Cs$(6^2P_{1/2})$ dissociation limit.}
\end{figure}

\begin{figure}
\centering
\includegraphics[width=0.4\textwidth]{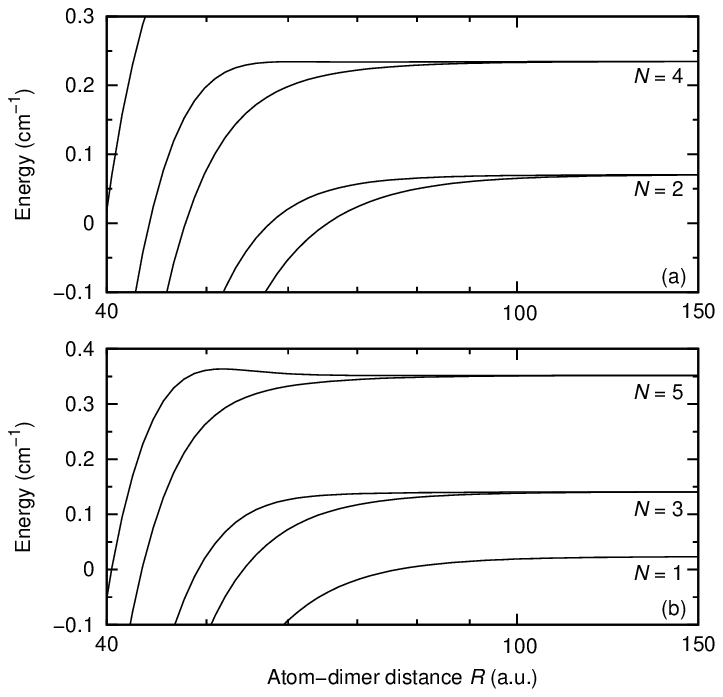}
\caption{\label{fig:Cs2-Cs-j12-Om32}Same as Fig. \ref{fig:Cs2-Cs-j12-Om12}, for $|\Omega|=3/2$, $j=1/2$.}
\end{figure}

\begin{figure}
\centering
\includegraphics[width=0.4\textwidth]{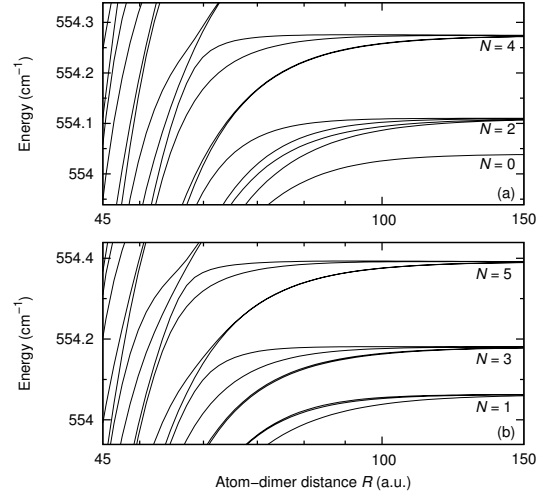}
\caption{\label{fig:Cs2-Cs-j32-Om12}Same as Fig. \ref{fig:Cs2-Cs-j12-Om12}, for $|\Omega|=1/2$, $j=3/2$.}
\end{figure}

\begin{figure}
\centering
\includegraphics[width=0.4\textwidth]{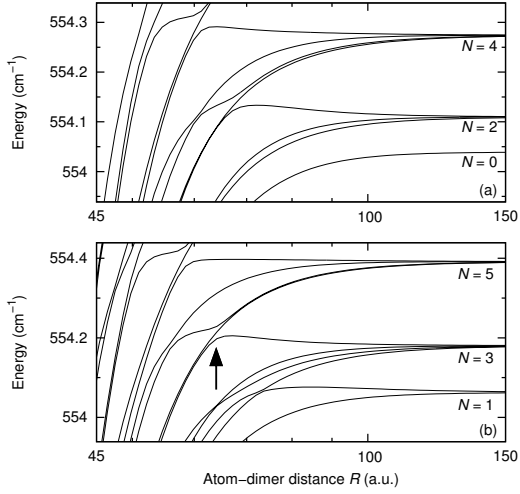}
\caption{\label{fig:Cs2-Cs-j32-Om32}Same as Fig. \ref{fig:Cs2-Cs-j12-Om12}, for $|\Omega|=3/2$, $j=3/2$. On panel (b), the arrow points out the three-curve crossings discussed in the text.}
\end{figure}

The potential energy curves shown on Figs.~\ref{fig:Cs2-Cs-j12-Om12} and \ref{fig:Cs2-Cs-j12-Om32} have very similar features. Most dimer rotational levels split into two curves as the two fragments get closer to each other. Typically, for $R>80$~a.u., one curve, characterized by a negative $C_6$ coefficient, is attractive, while the other, characterized by a positive $C_6$ coefficient, is repulsive. The  positive $C_6$ coefficients are due to the highly-negative parallel polarizability of the dimer at the atomic $6^2P_{1/2}\to 6^2S$ transition frequency (-5160~a.u.), compared to -3037~a.u.~at the $6^2P_{3/2}\to 6^2S$ transition frequency. For lower atom-dimer distances, the repulsive curves turn attractive, due to the coupling with the attractive curve connected to the higher dissociation limit. The resulting long-range potential barriers, whose height can go up to 0.1~cm$^{-1}$, could prevent collisions in the ultra-cold regime.
\begin{table}
\centering
\begin{tabular}{ccc}
~~$N$~~ & ~~$\left|\Omega\right|$~~ & $C_{6}$ (a.u.) \\
\hline
 0 & $1/2$ & -11022 \\
 1 & $1/2$ & -19952 \\
 1 & $1/2$ &   6840 \\
 1 & $3/2$ & -19952 \\
\end{tabular}
\caption{\label{tab:C6-j12} The $C_6$ coefficients of the Cs$_2(X^1\Sigma_g^+, v_d=0, N)$+Cs($6^2P_{1/2}$) long-range interaction calculated for $N=0$ and 1. In analogy to a diatomic molecule, the states are sorted by absolute values of the total angular momentum projection $\left|\Omega\right|$ on the $Z$ axis. All the $C_5$ coefficients are zero.}
\end{table}

For Cs in the $6^2P_{3/2}$ level, the potential energy curves (Figs.~\ref{fig:Cs2-Cs-j32-Om12} and \ref{fig:Cs2-Cs-j32-Om32}) look very different from the previous ones. Like those obtained in Paper III without spin-orbit interaction, most of them are attractive, and give birth to complex couplings for $R<80$~a.u.. Those couplings can consist of the two-curve crossings, like the ones described in Paper III, but also of three-curve crossings. The most striking example of such a crossing is pointed out by an arrow on Fig.~\ref{fig:Cs2-Cs-j32-Om32}(b), but this feature is quite general. They will be described in more details in the next paragraph, with Li$_2+$Li as a direct comparison with the curves in the underlying spinless symmetries will be possible.

The corresponding $C_5$ and $C_6$ coefficients are presented in Table \ref{tab:C5-C6-j32}, for $N=0$ and 1. The square radius of the $6p_{3/2}$ orbital of the cesium was calculated in our group with a Dirac-Fock method, which gives $\left\langle{r}_{6p_{3/2}}^2\right\rangle=78.45$~a.u.. The quadrupole moment of Cs$_2$ is the same as in Paper I ($q_2^0=18.56$a.u.). The $C_5$ and $C_6$ coefficients are of the same order of magnitude as those obtained in the LS coupling case (see Papers I and II). The most attractive curve, found for $N=1$ and $|\Omega|=1/2$, has inherited its behavior from the most attractive $\Sigma^+$ curve discussed in Paper I and II, although its $C_5$ coefficient is less attractive (-1131~a.u. with respect to -1674~a.u.). The associated $C_6$ coefficient is positive, but much lower the one in LS coupling case (2500~a.u. compared to 51249~a.u.). By contrast, all the other $C_6$ coefficients are negative.

\begin{table}
\centering
\begin{tabular}{ccccc}
~~$N$~~ & ~~$\left|\Omega\right|$~~ & $C_{5}$ ($\left\langle{r}_{6p_{3/2}}^2\right\rangle q_2^0$)
 & $C_{5}$ (a.u.) & $C_{6}$ (a.u.) \tabularnewline
\hline
 0 & $1/2$ & 0 & 0 & -50312 \tabularnewline
 0 & $3/2$ & 0 & 0 & -28580 \tabularnewline
 1 & $1/2$ & $-\frac{6}{25}\left(\sqrt{5}+1\right)$ & -1131 & 2500 \tabularnewline
 1 & $1/2$ & 0 & 0 & -35538 \tabularnewline
 1 & $1/2$ &  $\frac{6}{25}\left(\sqrt{5}-1\right)$ & 432 & -72712 \tabularnewline
 1 & $3/2$ & $\frac{9-\sqrt{105}}{25}$ &  -73 & -39759 \tabularnewline
 1 & $3/2$ & $\frac{9+\sqrt{105}}{25}$ & 1121 & -67219 \tabularnewline
 1 & $5/2$ & -$\frac{6}{25}$ & -349 & -23944 \tabularnewline
\end{tabular}
\caption{\label{tab:C5-C6-j32} The $C_{5}$ and $C_6$ coefficients of the Cs$_2(X^1\Sigma_g^+, v_d=0, N)$+Cs($6^2P_{3/2}$) long-range interaction calculated for $N=0$ and 1. In analogy to a diatomic molecule, the states are sorted by absolute values of the total angular momentum projection $\left|\Omega\right|$ on the $Z$ axis. As well as in Paper I, the values of $C_5$ are given in scaled units of $\left\langle{r}_{6p_{3/2}}^2\right\rangle q_2^0$.}
\end{table}

\section{Case "1A": Li$^*$+A$_{2}$}
\label{sec:Li-A2}

\begin{figure}
\centering
\includegraphics[width=0.4\textwidth]{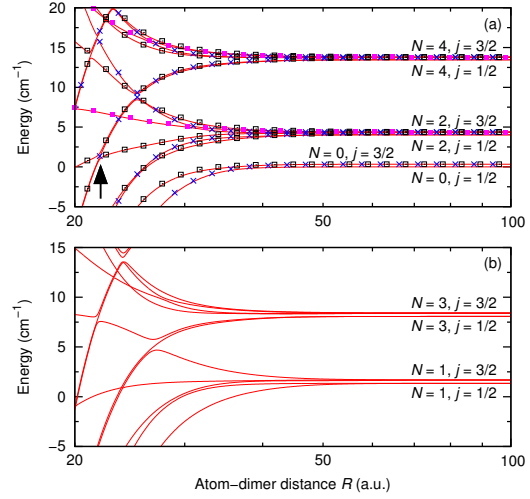}
\caption{\label{fig:Li2-Li-Om12}(Color online) Long-range potential energy curves (without dipolar interaction) between a ground-state Li$_{2}$ and an excited Li($2^{2}P$) atom, as functions of their mutual separation $R$, for $|\Omega|=1/2$ and for: (a) the even values of $N$ and (b) the odd values of $N$. On panel (a), potential curves calculated by neglecting the fine structure of Li($2^{2}P$) are also displayed: in the $\Sigma^{+}$ (crosses), $\Sigma^{-}$ (full squares) and $\Pi$ symmetries (empty squares). The arrow points out the three-curve crossings discussed in the text.}
\end{figure}

Now the two atomic fine-structure levels are so close to each other that they can be coupled by the electrostatic interaction. This is illustrated by a lithium atom in the $2^2P$ state and both by Li$_2$ and K$_2$ dimers. The results presented in this section are obtained by diagonalizing $\hat{W}_2$ (see Eq.~(\ref{eq:W2})), but without the second-order dipole-dipole interaction. The square radius of the $2p$ orbital of the lithium atom was calculated in our group with an Hartree-Fock method, which gives $\left\langle{r}_{2p}^2\right\rangle=25.50$~a.u.. The quadrupole moments of Li$_2$ and K$_2$ are equal to 10.73 and 15.69 a.u., respectively\cite{harrison2005}. First, the potential energy curves between Li$_2$ and Li are displayed on Fig.~\ref{fig:Li2-Li-Om12} for $|\Omega|=1/2$. On panel (a), for the even values of the rotational quantum number $N$, the curves of symmetries $\Sigma^\pm$ and $\Pi$, calculated without the atomic fine structure, are also plotted. As the splitting between two subsequent rotational levels of Li$_2$, at least $6B_0\approx4$~cm$^{-1}$, is much higher that the fine-structure splitting (0.3~cm$^{-1}$), the description in LS coupling case seems adequate, especially in the coupling region, for $20<R<40$~a.u., where each curve belonging to the $|\Omega|=1/2$ can be clearly identified with a curve belonging either  to the $\Sigma^\pm$ or to the $\Pi$ symmetry. We can also see three-curve crossings (see arrow on panel (a)), like in the case Cs$_2+$Cs. This crossing concerns one $\Sigma^+$ and two $\Pi$ states. Outside the crossing, one of the $\Sigma^+$ states and the $\Pi$ state are almost degenerate, whereas, at the crossing, the two $\Pi$ states avoid each other. The mechanism is the same for cesium, even if the curves resulting from the calculation in the $jj$ coupling case do not have such strong components coming from \textit{one} given LS-coupling-case symmetry.

\begin{figure}
\centering
\includegraphics[width=0.4\textwidth]{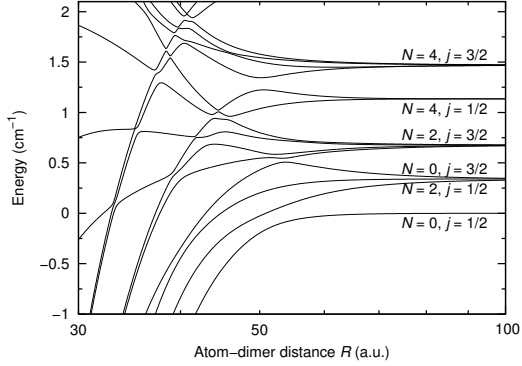}
\caption{\label{fig:K2-Li-Om12}Long-range potential energy curves (without dipolar interaction) between a ground-state K$_{2}$ and an excited Li($2^{2}P$) atom, as functions of their mutual separation $R$, for $|\Omega|=1/2$ and the even values of $N$.}
\end{figure}

At last, we consider the interaction between Li($2^2P$) and K$_2$ for which potential energy curves are plotted on Fig.~\ref{fig:K2-Li-Om12} for $|\Omega|=1/2$. This situation is particularly interesting since two asymptotic channels namely $N=0,j=3/2$ and $N=2,j=1/2$ respectively at 0.335 cm$^{-1}$ and 0.340 cm$^{-1}$, are almost degenerate. The two channels can thus be coupled at large distances: for instance at $R=120$~a.u., the three curves connected to those two limits and taken with increasing energies decompose as (0.2;0.8), (1;0), and (0.8;0.2) on the channels ($N=0,j=3/2$; $N=2,j=1/2$). Moreover, the coupling region for $30<R<70$~a.u. is characterized by numerous avoided crossings.

\section{Discussion}
\label{sec:discussion}

The results displayed in the preceding section clearly show that a complicated dynamics is likely to occur at large distances, namely well beyond the range of chemical forces,  during the cold collision between an excited atom and a ground state molecule. This situation represent a new possibility to investigate the range of validity of the universal model for inelastic collisions at ultracold energies \cite{julienne2009,idziaszek2010a}. Based on Multichannel Quantum Defect Theory (MQDT), this model considers that the rate for an ultracold inelastic collision proceeds first from the long-range interactions which controls the probability for the system to reach the range of chemical forces at short distances. A long-range potential with a dominant term $C_n/R^n$ is characterized by an interaction length $\bar{a}(n)$ \cite{gribakin1993,julienne2009}
\begin{equation}
\bar{a}(n)=\cos\left(\frac{\pi}{n-2}\right)
\left(\frac{2\mu C_n}{\hbar^2 (n-2)^2}\right)^{\frac{1}{n-2}}
\frac{\Gamma\left(\frac{n-3}{n-2}\right)}{\Gamma\left(\frac{n-1}{n-2}\right)}
\label{eq:a_of_n}
\end{equation}
where $\mu$ is the reduced mass of the system, and $\Gamma$ the usual Gamma function. A characteristic energy $\bar{E}(n)=\hbar^2/(2\mu \bar{a}(n)^2)$ is associated to this length. 
For collision energies $E=\hbar^2 k^2/2\mu$ dominated by $s$-wave scattering, \textit{i.e.} $k\bar{a}(n) << 1$, the inelastic rate is independent of $k$ and writes $K^{in}(n)=2 h \bar{a}(n)/\mu$ \cite{julienne2009}. In the present case, the dominant term for the interaction between a ground state Cs$_2$ molecule and a Cs($P_{3/2}$) atom corresponds to $n=5$, and for a typical value $C_5=-1131$~a.u. (see Table \ref{tab:C5-C6-j32}), one obtains $\bar{a}(5)=0.364506 (2 \mu C_5/\hbar^2)^{1/3} \approx 260$~a.u.. For a Cs($P_{1/2}$) atom ($n=6$) one has $\bar{a}(6)=0.477989(2\mu C_6/\hbar^2)^{1/4} \approx 103$~a.u.. In both case, the characteristic length is larger than the intermediate range of distances identified above where spin-orbit coupling, rotational energy and electrostatic energy all compete together. Therefore such an ultracold inelastic collision will be actually defined by three domains: (i) the large distances where the sole $C_n/R^n$ term controls the scattering, (ii) the intermediate distances above with a specific treatment of the dynamics, (iii) the short distances where one can reasonably assume that so many channels are open that the reaction probability is equal to unity. The dynamics in the intermediate range could well be treated within a coupled-channel framework in one dimension as implemented in Ref.\cite{hudson2008} for ultracold ground state Rb (or Cs) atoms and RbCs molecules. Thus it is most likely that the validity of the universal model of Refs.\cite{julienne2009,idziaszek2010a} will be limited in the present case, just like it has been argued for ultracold collisions between ground state KRb molecules in their lowest rovibrational levels \cite{ospelkaus2010a}.

Another remarkable expected feature of such collisions will be the existence of numerous Feshbach resonances induced by bound levels of the trimer close to one dissociation limit Cs$*$+Cs$_2$($N$) interacting with the continuum related to a dissociation limit with a smaller value of $N$. Such resonances may enhance the photoassociation probability and the decay down to stable molecules through the $R$-transfer of the probability density of the system to smaller distances, as it is well known for atom-atom photoassociation\cite{vatasescu2000,dion2001,viteau2008}. 

Finally, it is also worthwhile to mention that the energy spacing between molecular rotational levels is -at least for Cs$_2$- of the same order of magnitude than the hyperfine splitting of the Cs ground state, which could again induce a complex resonant dynamics at large distances, which will the subject of a further study.  

\section*{Acknowledgements}
Stimulating discussions with N. Bouloufa, V. Kokoouline, and R. Vexiau, are gratefully acknowledged. M.L. acknowledges the support of Triangle de la Physique in the framework of the contract QCCM-2008-007T Quantum Control of Cold Molecules.




\footnotesize{
\bibliography{bibliocold} 
\bibliographystyle{rsc} 
}

\end{document}